\documentclass[%
 reprint,superscriptaddress,amsmath,amssymb,prb]{revtex4-1}

\usepackage{graphicx}
\usepackage{dcolumn}
\usepackage{bm}

\begin{document}

\title{Kondo behavior and metamagnetic phase transition in a heavy fermion compound CeBi$_2$}

\author{W. Zhou}
\affiliation{Advanced Functional Materials Lab and Department of Physics, Changshu Institute of Technology, Changshu 215500, China}
\author{C. Q. Xu}
\affiliation{Advanced Functional Materials Lab and Department of Physics, Changshu Institute of Technology, Changshu 215500, China}
\author{B. Li}
\affiliation{Information Physics Research Center, Nanjing University of Posts and Telecommunications, Nanjing, 210023, China}
\author{R. Sankar}
\affiliation{Center for Condensed Matter Sciences, National Taiwan University, Taipei 10617, Taiwan}
\author{F. M. Zhang}
\affiliation{Advanced Functional Materials Lab and Department of Physics, Changshu Institute of Technology, Changshu 215500, China}
\author{B. Qian}
\email{njqb@cslg.edu.cn}
\affiliation{Advanced Functional Materials Lab and Department of Physics, Changshu Institute of Technology, Changshu 215500, China}
\author{C. Cao}
\affiliation{Department of Physics, Hangzhou Normal University, Hangzhou 310036, China}
\author{J. H. Dai}
\affiliation{Department of Physics, Hangzhou Normal University, Hangzhou 310036, China}
\author{Jianming Lu}
\affiliation{High Field Magnet Laboratory (HFML-EMFL), Radboud University, Toernooiveld 7, 6525ED Nijmegen, Netherlands}
\author{Xiaofeng Xu}
\email{xiaofeng.xu@cslg.edu.cn}
\affiliation{Advanced Functional Materials Lab and Department of Physics, Changshu Institute of Technology, Changshu 215500, China}

\date{\today}

\begin{abstract}
Heavy fermions represent an archetypal example of strongly correlated electron systems which, due to entanglement among different interactions, often exhibit exotic and fascinating physics involving Kondo screening, magnetism and unconventional superconductivity. Here we report a comprehensive study on the transport and thermodynamic properties of a cerium-based heavy fermion compound CeBi$_2$ which undergoes an anti-ferromagnetic transition at $T_N$ $\sim$ 3.3 K. Its high temperature paramagnetic state is characterized by an enhanced heat capacity with Sommerfeld coefficient $\gamma$ over 200 mJ/molK$^2$. The magnetization in the magnetically ordered state features a metamagnetic transition. Remarkably, a large negative magnetoresistance associated with the magnetism was observed in a wide temperature and field-angle range. Collectively, CeBi$_2$ may serve as an intriguing system to study the interplay between $f$ electrons and the itinerant Fermi sea.
\end{abstract}

\maketitle

\section{Introduction}
The interactions between local moments and itinerant electrons may result in many exotic quantum phenomena, the heavy fermions in intermetallic lanthanide/actinide compounds being a clear-cut example\cite{Stewart}. In heavy fermions, the lattices of localized $f$-moments are antiferromagnetically coupled with conduction electrons via Kondo interaction and below a characteristic temperature, referred to as the Kondo coherence temperature, a Landau Fermi liquid with significantly enhanced electron effective mass emerges. Besides the Kondo interaction, the competing Ruderman-Kittel-Kasuya-Yosida (RKKY) interaction favors the magnetic ordering of the $f$-moments\cite{LuXin,Analytis}.

Among heavy fermions, Ce-based compounds stand for the simplest example that incarnates most of the important physics since each Ce$^{3+}$ ion has only one $f$-electron in the 4$f$ shell\cite{Budko,Petrovic,Bauer,Yuan-npj,Luo-npj}. Specifically, many Ce-based antimonides/bismuthides display intriguing properties at low temperatures, including complex magnetic ordering, heavy-fermion behavior, unconventional superconductivity, or even nontrivial topological states\cite{Petrovic,Yuan-npj,Petrovic-JMMM,Tafti}. Taking the rare-earth monoantimonide CeSb as an example, its magnetic resistivity logarithmically increases with decreasing $T$ due to the incoherent Kondo scattering and at least four metamagnetic transitions are observed in the magnetization curves\cite{Analytis,Yuan-npj}. More intriguingly, the angle-dependent magnetoresistance shows the signature of nontrivial Weyl fermions in its ferromagnetic state, which makes CeSb a promising candidate in the nascent search for novel topological states in the strongly correlated regime\cite{Yuan-npj}.

Another cerium-based intermetallic system, as yet much less studied, is the cerium-dibismuthide. A previous work focusing on the magnetization properties of CeBi$_2$ has reported the discovery of a metamagnetic phase transition\cite{Petrovic-JMMM}. However, similar to the Weyl semimetal candidate CeSb, the physical properties of the magnetic CeBi$_2$ deserve more fundamental exploration. In particular, it is not yet clear how universal the topological Weyl point seen in CeSb is in other Ce-based antimonides/bismuthides. In this context, we performed a systematic study on the transport and thermodynamic properties of CeBi$_2$ single crystals. Our study reveals that CeBi$_2$ is a typical Kondo lattice system with a large Sommerfeld coefficient $\gamma$ over 200 mJ/mol K$^2$. Contrasting to the topological origin of the negative magnetoresistance (MR) in CeSb\cite{Yuan-npj}, the negative MR in CeBi$_2$ can be instead associated with the complex magnetism of the system.

\section{Experiment}

Single crystals of CeBi$_2$ were grown by the self-flux method. Elemental Bi and Ce were mixed with a molar ratio over 10:1. The mixture was then sealed in a quartz tube in vacuum. The tube was heated to 1000$^\circ$C in a high-temperature box furnace and kept at this temperature for several hours. After a slow cooling process, plate-like crystals which can be easily exfoliated were harvested.

Single crystal X-ray diffraction (XRD) measurements were performed at room temperature using a Rigaku diffractometer with Cu $K$$\alpha$ radiation and a graphite monochromator. Low-field MR, angle-dependent MR, and specific heat data were all collected on a Quantum Design (QD) physical property measurement system (PPMS). The MR data in high magnetic field were acquired in a resistive Bitter magnet at HFML (Nijmegen), with maximum field 35 T. The samples displayed highly symmetric resistance signals with respect to the magnetic field direction, hence only data obtained for positive magnetic field orientation are shown. The magnetization data were measured using a QD superconducting quantum interference device magnetometer (SQUID) with fields up to 7 T.

\section{Results}

\begin{figure}
\includegraphics[width=9cm]{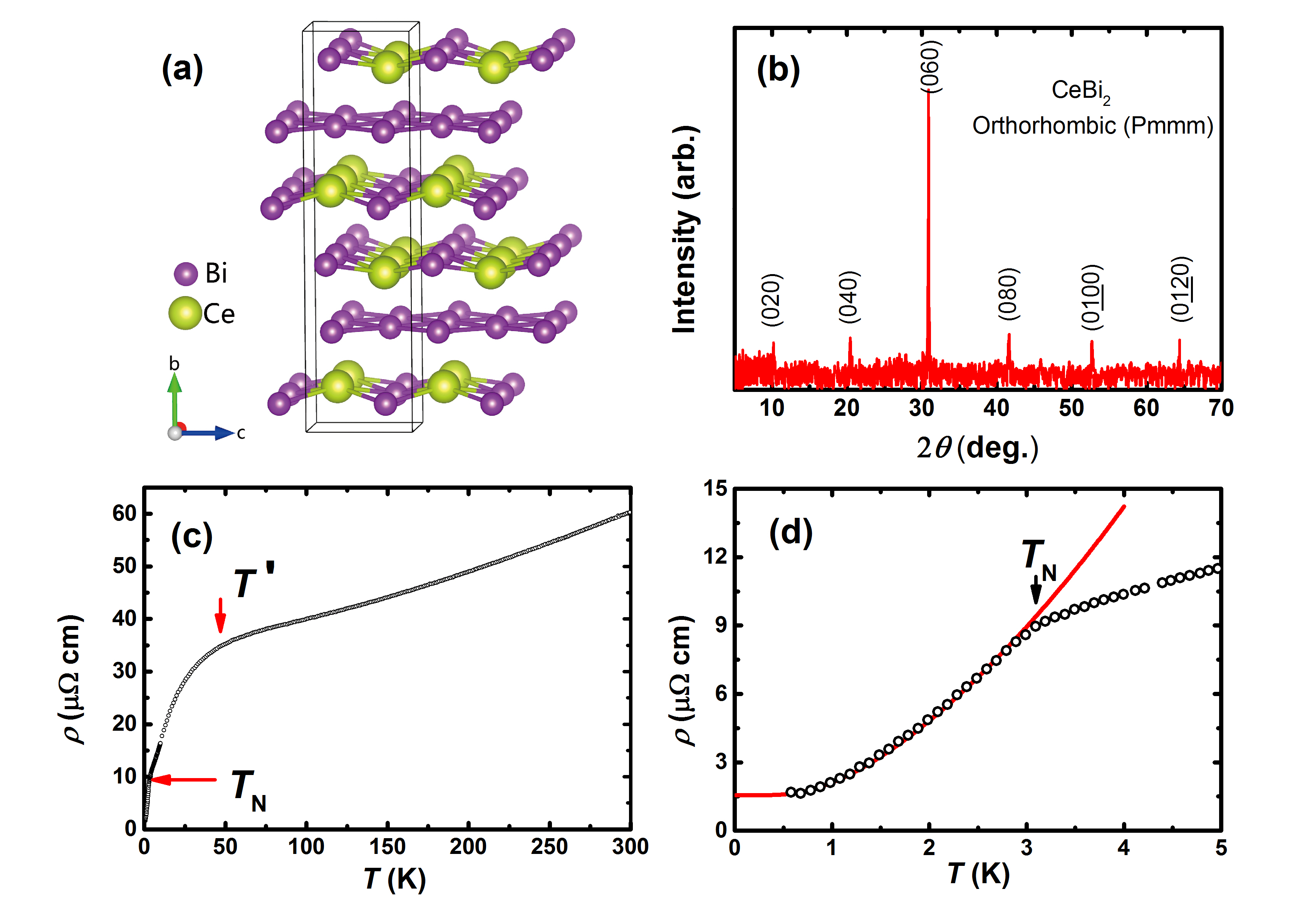}
\caption{\label{RT} (a) The crystal structure of CeBi$_2$. The single layers of Bi and the CeBi bilayers are alternately stacked along the $b$-axis. The solid box indicates one conventional unit cell. (b) The X-ray diffraction pattern for a CeBi$_2$ single crystal. (c) Temperature dependence of resistivity $\rho$ of CeBi$_2$. $T'$ and $T_N$ denote two characteristic temperatures where $\rho$ shows marked decreases with $T$. (d) A blow-up of the low-$T$ resistivity in panel (c). The red line is the fit to Eqn. \ref{ResAFM}.}
\label{res}
\end{figure}

As shown in Fig. \ref{res} (a), the atomic structure of CeBi$_2$ is layered, with single layers of Bi separated from each other by CeBi bilayers and stacked along the crystalline $b$-axis. The Bi atoms in the Bi single layer are squarely coordinated while in the CeBi bilayers, Ce atoms resides at the center of the square formed by four Bi atoms. A typical XRD pattern is shown in Fig. \ref{res} (b). Only the (0\underline{2$\ell$}0) peaks were observable, suggesting that the crystallographic $b$ axis is perfectly perpendicular to the surface facet of the crystal. All (0\underline{2$\ell$}0) peaks can be well indexed with the orthorhombic (Pmmm, No. 47) space group, sharing the same structure as LaBi$_2$. The crystal lattice parameter $b$ is estimated as 17.37{\AA}. Fig. \ref{res} (c) depicts the $T$ dependence of the electrical resistivity $\rho(T)$ for a CeBi$_2$ single crystal with current flowing along the $ac$-plane. The resistivity of CeBi$_2$ is characterized by two sharp drops at temperatures $T'$ and $T_N$. The resistivity drop at $T'$ may be attributed either to the coherence in Kondo scattering or to the crystal electric-field (CEF) splitting of Ce atoms\cite{YbPtBi}. As will be seen below, the characteristic $T_N$ can be assigned to an antiferromagnetic (AFM) phase transition. In the AFM state, the resistivity drops dramatically due to the reduction in electron-spin scattering in the ordered state. In this case, the resistivity takes the form\cite{Pikul-CeNiGe3,CeNiSi,FontesPRB,FontesPhysicaB}:

\begin{equation}
\rho(T)=\rho_0+A_\rho\Delta^2\sqrt{\frac{T}{\Delta}}e^{-\Delta/T}[1+\frac{2}{3}(\frac{T}{\Delta})+\frac{2}{15}(\frac{T}{\Delta})^2],\label{ResAFM}
\end{equation}

\noindent where $\Delta$ is the gap size of the spin-wave spectrum in the AFM state. As seen from Fig. \ref{res}(d), the resistivity of CeBi$_2$ in the AFM state fits well to the above formula, yielding $\rho_0$=1.56 $\mu\Omega$cm, $A_\rho$=1.25 $\mu\Omega$cm/K$^2$, and $\Delta$=2.4 K.

\begin{figure}
\includegraphics[width=8.5cm]{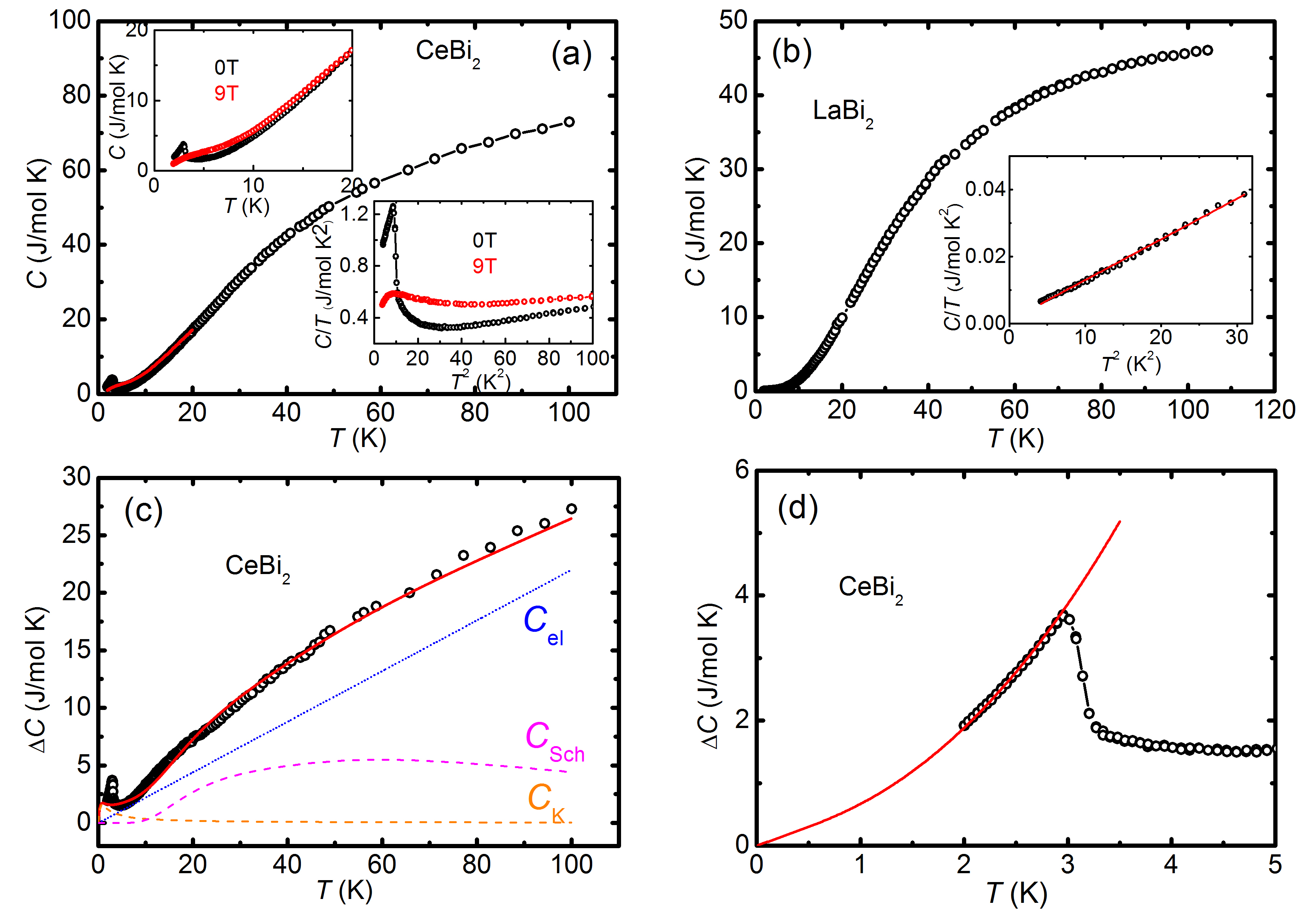}
\caption{\label{RT} (a) Temperature dependence of heat capacity $C$ of CeBi$_2$. The red line is the heat capacity under a 9 T field ($H$$\parallel$$b$). The upper inset shows the low-$T$ heat capacity below 20 K. The lower inset manifests the data as $C/T$ vs $T^2$ below 10K. (b) Temperature dependence of heat capacity of the reference compound LaBi$_2$. The inset is the low-$T$ plot of $C/T$ vs $T^2$ to separate the electronic and phononic contributions. (c) The non-phononic heat capacity $\Delta$$C$ of CeBi$_2$. The thick red line is the fit to $\Delta$$C$ in the paramagnetic state, assuming the nonlattice contributions come from electron ($C_{el}$, Kondo ($C_K$) and CEF effect ($C_{Sch}$), each of which has also been plotted as individual lines. (d) Zoom-in of the low-$T$ $\Delta$$C$ in the AFM state. The red line is a fit to Eqn. \ref{HCAFM}.}
\label{HC}
\end{figure}

\begin{figure*}
\includegraphics[width=15cm]{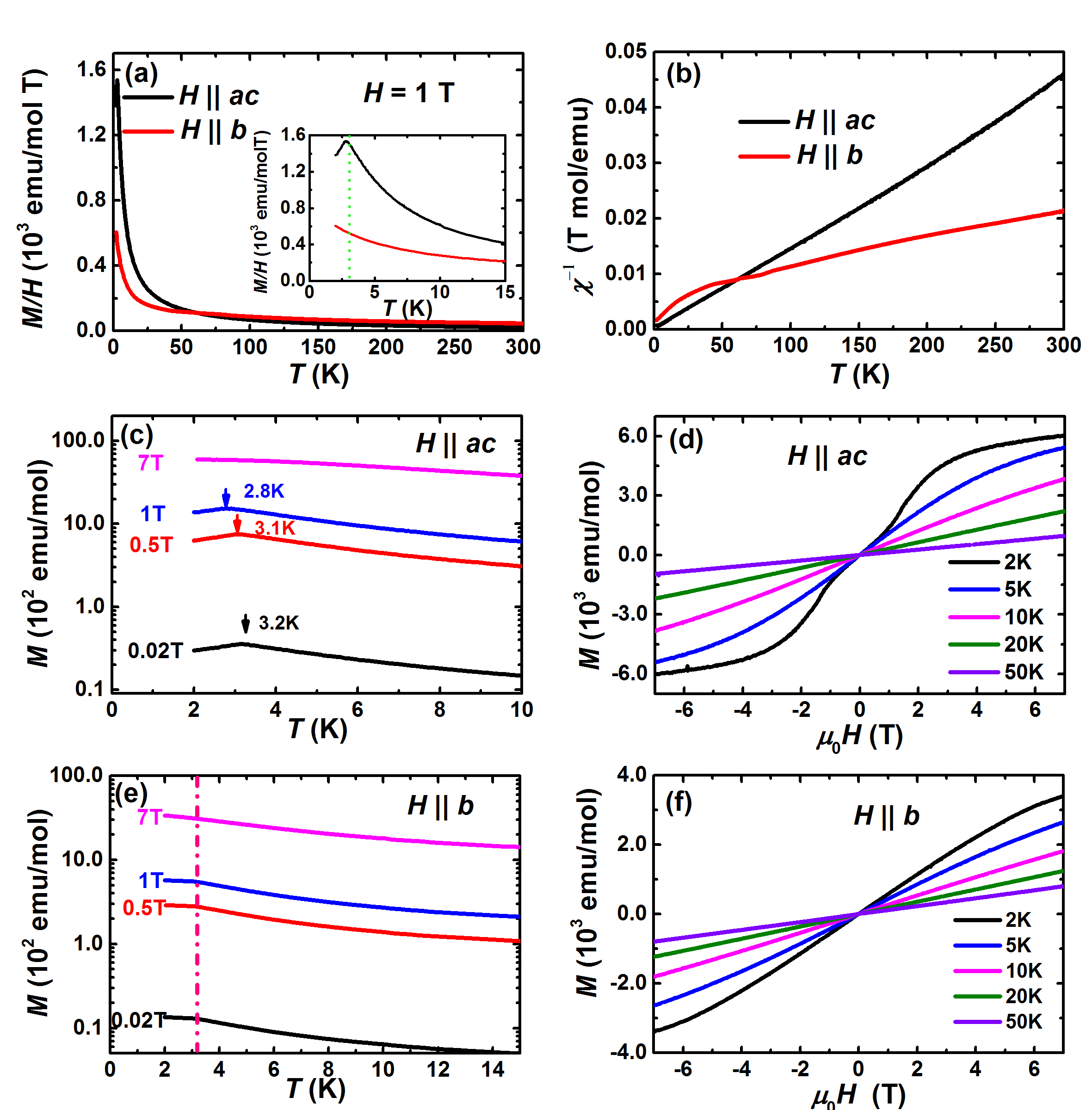}
\caption{\label{RT} (a) Temperature dependence of molar magnetic susceptibility of single crystalline CeBi$_2$ measured under a field of 1 Tesla oriented perpendicular and parallel to the $b$-axis. (b) Inverse magnetic susceptibility as a function of temperature to demonstrate the Curi-Weiss law. (c) Magnetization under different fields ($H$$\parallel$$ac$). (d) Isothermal magnetization curves at some fixed temperatures for $H$$\parallel$$ac$. (c) Magnetization under different fields ($H$$\parallel$$b$). (d) Isothermal magnetization curves at some fixed temperatures for $H$$\parallel$$b$. }
\label{Mag}
\end{figure*}

The specific heat of CeBi$_2$ is shown in Fig. \ref{HC} (a). Under zero field, a sizeable jump at $T_N$ is observed, consistent with a second-order phase transition. Under a field of $H$ = 9 T ($\parallel$ $b$-axis), the jump is greatly suppressed, while the transition temperature shows no clear shift. At $T$$\sim$20 K, the field has little effect on the heat capacity and two curves overlap (the upper inset of Fig. \ref{HC} (a)). The entropy under 9 T is approximately the same as that in 0 T at 20 K. The effect of the magnetic field on the heat capacity is better seen when we plot $\frac{C}{T}$ as a function of $T^2$ (the lower inset of Fig. \ref{HC} (a)). In this manner, the electronic specific heat coefficient $\gamma$ can be estimated crudely as $\gamma$$\sim$ 200 mJ/mol K$^2$ if we linearly extrapolate the data from above $\sim$ 6 K. The detailed analysis of heat capacity however will be given below.

\begin{figure*}
\includegraphics[width=16cm]{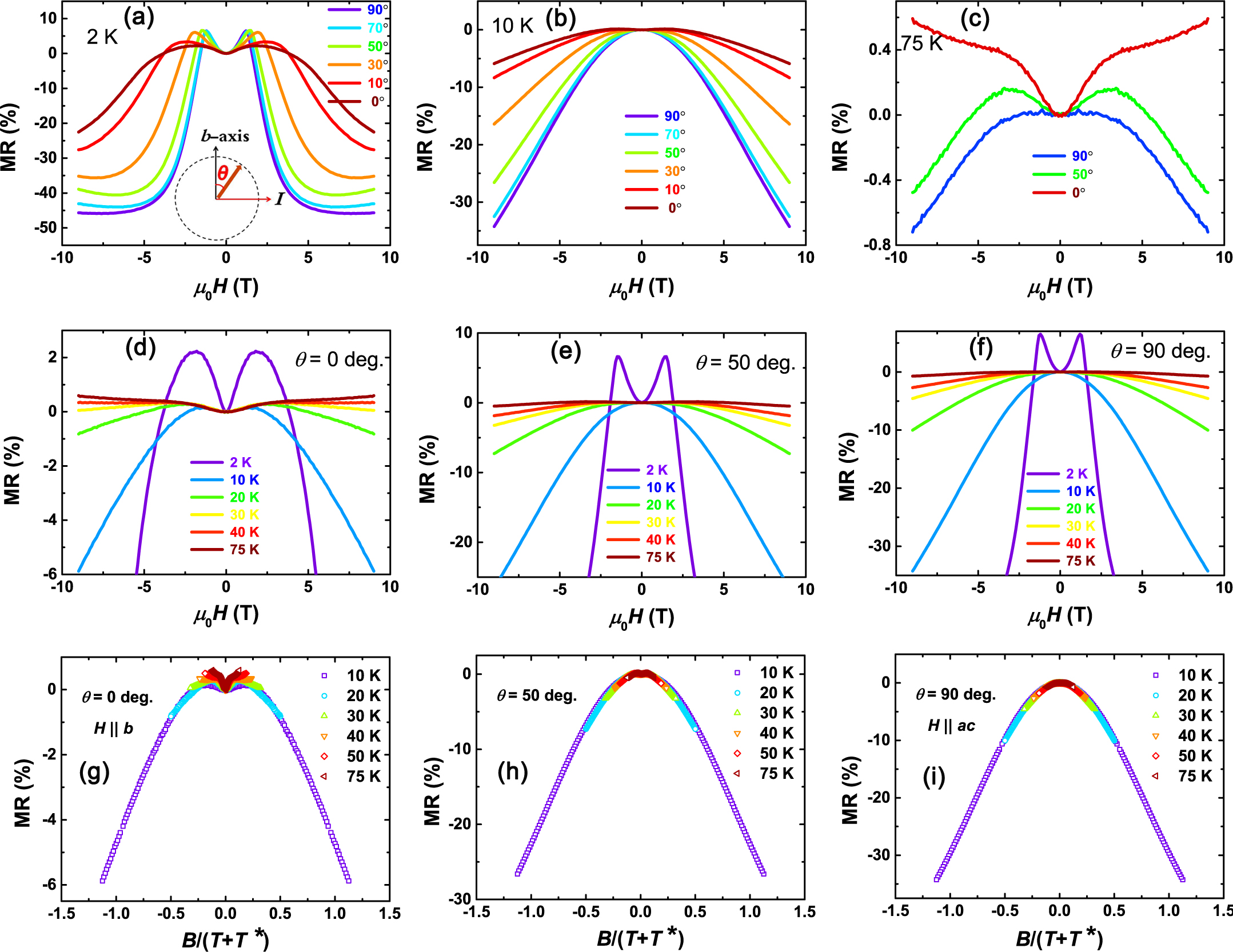}
\caption{\label{RT} (a-c) Field dependence of MR measured at various angles for three representative temperatures. Inset of panel (a) shows the field configurations. The current was set in the $ac$ plane and the field was confined in the plane constructed by the crystal $b$-axis and the current flowing direction. $\theta$ is defined as the angle between the field direction and the crystal $b$-axis. (d-f) MR measured at different temperatures for three representative angles. (g-i) The corresponding Schlottmann's scaling of the isothermal MR. $T^*$ is the characteristic scaling temperature.}
\label{MR}
\end{figure*}

To determine the electronic Sommerfeld coefficient $\gamma$ of the sample, we refer to the specific heat of the nonmagnetic counterpart LaBi$_2$ crystal which is isostructural to CeBi$_2$. The phonon contribution to the heat capacity in LaBi$_2$ can be obtained by subtracting the electronic term which is determined by fitting the low-$T$ heat capacity to $C(T)$=$\gamma$$T$+$\beta$$T^3$, as shown in the inset of Fig. \ref{HC} (b). Assuming the same phononic heat capacity in CeBi$_2$ and LaBi$_2$, we can get the non-phononic heat capacity $\Delta$$C$ of CeBi$_2$, as plotted in Fig. \ref{HC} (c). In the paramagnetic region of a Kondo lattice, $\Delta$$C$ is comprised of three components, i.e., $\Delta$$C$=$C_{el}$+$C_K$+$C_{Sch}$, where $C_{el}$, $C_K$, $C_{Sch}$ represents the electronic, Kondo, Schottky terms respectively\cite{Pikul-CeNiGe3}. The Kondo term $C_K$ in a Kondo lattice is only dependent on the Kondo temperature $T_K$ and was derived theoretically in the original paper by Schotte\cite{Schotte1,Schotte2}. The Schottky anomaly with energy gaps of $\Delta_1$ and $\Delta_2$, resulting from the Ce$^{3+}$ crystal electric-field splitting, was considered\cite{Pikul-CeNiGe3}. Overall, these three terms can fit $\Delta$$C$ fairly well above $T_N$ as demonstrated by the thick red line in Fig. \ref{HC}(c) and the corresponding fitting parameters are $\gamma$=220 mJ/mol K$^2$, $\Delta_1$=75 K, $\Delta_2$=190 K and $T_K$=1.5 K. Their individual lines are also delineated in the figure. In the AFM state ($T<T_N$), however, $\Delta$$C$ has the form\cite{FontesPRB,ContinentinoPRB}:

\begin{equation}
\Delta C=\gamma_m T+A_C\Delta^4\sqrt{\frac{T}{\Delta}}e^{-\Delta/T}[1+\frac{39}{20}(\frac{T}{\Delta})+\frac{51}{32}(\frac{T}{\Delta})^2], \label{HCAFM}
\end{equation}

\noindent where $\gamma_m$ is the Sommerfeld coefficient in the magnetically ordered state and $\Delta$ is again the gap in the spin-wave spectrum (see Eqn. \ref{ResAFM}). Although the fit range is limited, it gives $\Delta$=2.9 K which is very close to the one obtained from the resistivity fit in Fig. \ref{res} (d). $\gamma_m$ and $A_C$ from the fit are 600 mJ/mol K$^2$ and 16 mJ/mol K$^5$ respectively. Note that this $\gamma_m$ is significantly enhanced compared with the one in the paramagnetic state, $\gamma$$\sim$220 mJ/mol K$^2$. Similar behaviors have been observed in other antiferromagnetic Kondo systems\cite{Pikul-CeNiGe3,CeNiSi}.

\begin{figure}
\includegraphics[width=9.8cm]{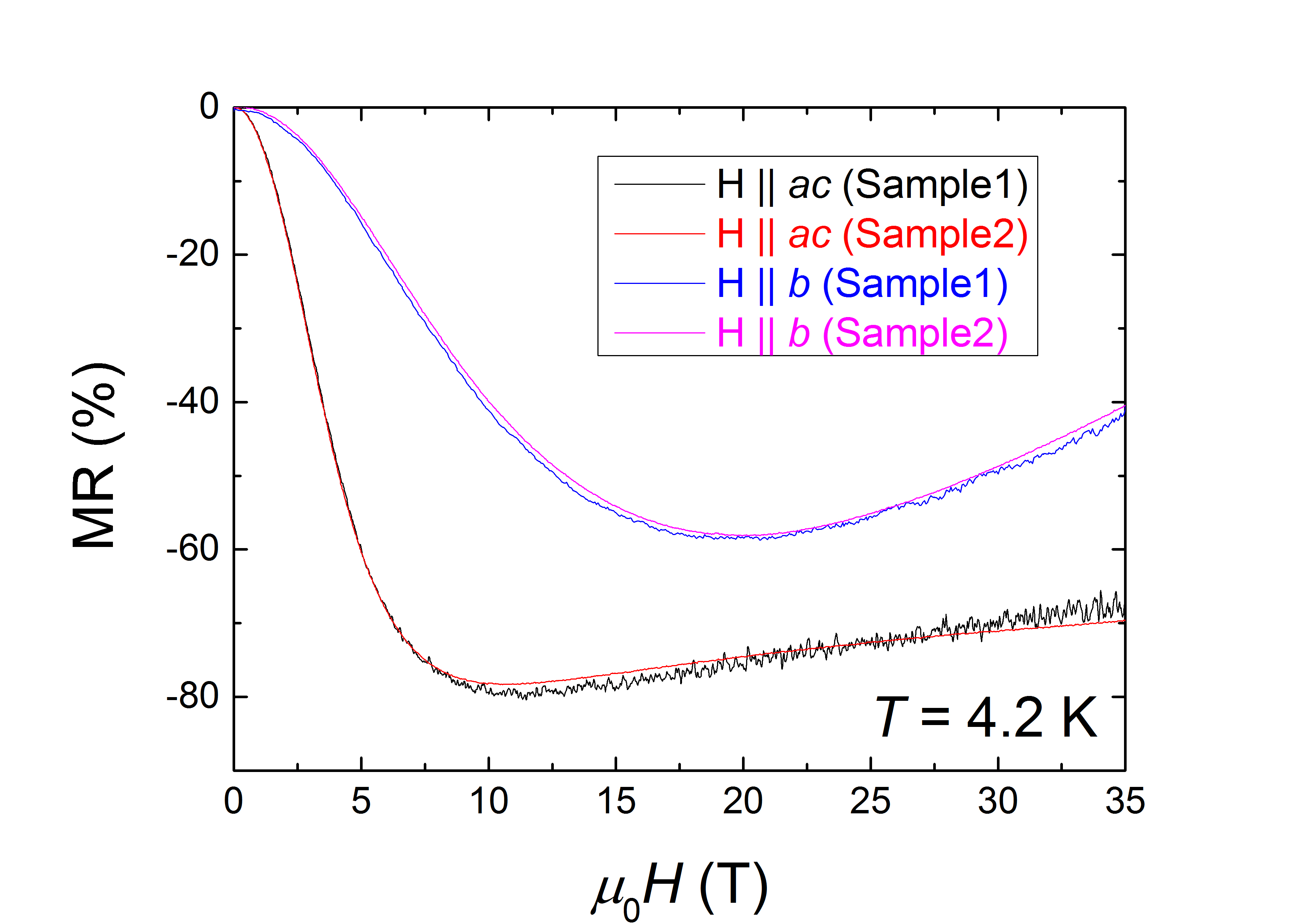}
\caption{\label{RT} MR at $T$=4.2 K with fields up to 35 T for two single crystals (labeled as $S1$ and $S2$).}
\label{highfield}
\end{figure}

Fig. \ref{Mag} summarizes the magnetization measurements on CeBi$_2$ for both field orientations, i.e., $H$$\parallel$$ac$-plane and $H$$\parallel$$b$-axis. Notable magnetization anisotropy is observed. As seen from the temperature dependence of the magnetic susceptibility ($M$/$H$) in Fig. \ref{Mag} (a), the susceptibility increases quickly below $\sim$50 K for both field directions. At $T$ $\sim$ $T_N$ (3.3 K), a pronounced peak in the $M$/$H$ curve with $H$$\parallel$$ac$ is seen while for $H$$\parallel$$b$, only a weak inflexion point is observed, as shown in the inset of Fig. \ref{Mag} (a), suggesting an antiferromagnetic phase transition. This anisotropic magnetization implies that the spins are aligned predominantly within the $ac$ plane\cite{XuPr124}.

To further analyze the susceptibility, the data are plotted in the panel (b) of Fig. \ref{Mag} as $\chi^{-1}$ versus temperature to compare with the Curie-Weiss law. As seen, the Curie-Weiss law is roughly obeyed for $H$$\parallel$$ac$ while for $H$$\parallel$$b$, deviation is seen below $\sim$50 K, presumably due to the CEF effects. We further extract the effective magnetic moment $\mu_{\rm eff}$ of Ce ions, using the Curie-Weiss law: $\chi$($T$)=$\frac{\mu_0 n \mu^2_{\rm eff}}{3k_B(T-\theta)}$, where $n$ is the number of magnetic ions per mole and $\theta$ is a parameter related to the interaction between the ions. The fits give $\mu_{\rm eff}$=2.3$\mu_B$, $\theta$= -2 K for $H$$\parallel$$ac$ and $\mu_{\rm eff}$=4.0$\mu_B$, $\theta$= -109 K for $H$$\parallel$$b$. The value for $H$$\parallel$$ac$ is consistent with the theoretical value of 2.54$\mu_B$ for a free Ce$^{3+}$ ion while the latter is somewhat exaggerated.

Fig. \ref{Mag} (c) shows the magnetization curves under different field strengths for $H$$\parallel$$ac$. As seen, the kink associated with the AFM transition is gradually suppressed to lower temperatures with increasing field and for $\mu_0H$ $>$ 2 T, this kink disappears and a ferromagnetic-like magnetization saturation at low temperatures emerges. This ferromagnetic-like saturation is more notable on the isothermal magnetization curves in low temperature regions (Fig. \ref{Mag} (d)). Besides, when $T$ $<$$T_N$, the $M$($H$) curves display a meta-magnetic phase transition at field around 1.4 T, revealing a complex magnetic structure of CeBi$_2$. At 2 K and 7 T, the magnetization reaches a value of 6000 emu/mol that corresponds to the magnetic moment of 1.1$\mu_B$, much lower than the saturated moment of Ce$^{3+}$ ion ($\mu_J$$\sim$2.14$\mu_B$). Hence, more metamagnetic transitions are likely in higher fields. Each of the $M$($H$) curves was performed with both increasing and decreasing field strength, yet no magnetic hysteresis was observed. By contrast, for $H$ $\parallel$ $b$, the magnetic properties become very different. Firstly, the AFM transition temperature $T_N$ show no clear shift with increasing field strength (Fig. \ref{Mag} (e)). Secondly, no obvious metamagnetic phase transition can be observed in the magnetization curves for $H$ $\parallel$ $b$ (Fig. \ref{Mag} (f)), indicating a rather strong magnetic anisotropy.

\begin{figure}
\includegraphics[width=8cm]{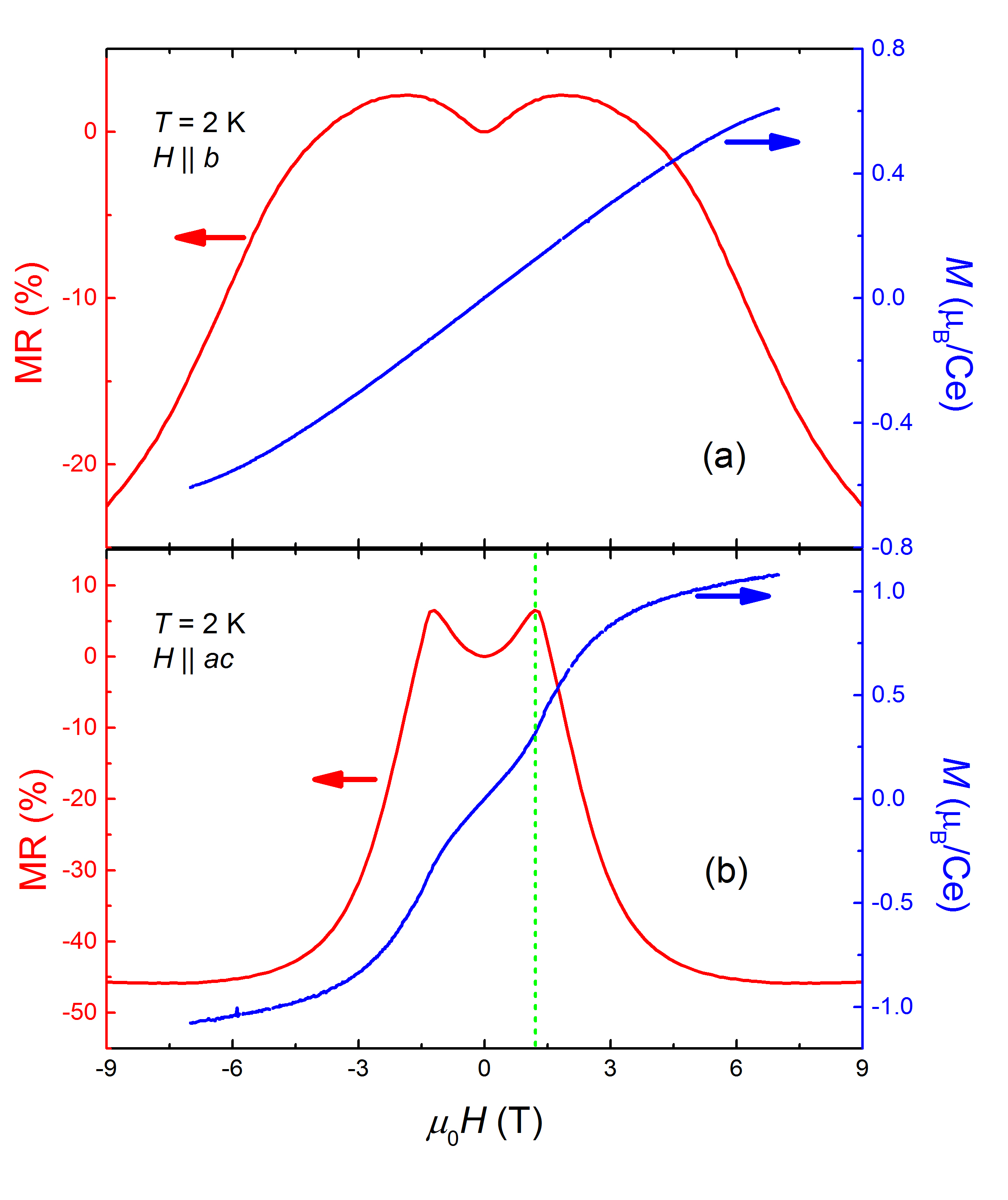}
\caption{\label{RT} Comparison between MR and magnetization for two field orientations at $T$ = 2 K.}
\label{comparefig}
\end{figure}

To explore the transport properties of CeBi$_2$, we performed the detailed magneto-transport measurements of CeBi$_2$ crystals at different temperatures and field angles. The field dependence of the magneto-resistance (MR), defined as MR=$\frac{\rho(H)-\rho(0)}{\rho(0)}$$\times$100$\%$, at different angles and for three representative temperatures is shown in Figs.\ref{MR} (a-c). The inset of Fig. \ref{MR} (a) sketches the configuration of current flow (along the $ac$-plane) and the field orientation. The magnetic field is constrained in the plane constructed by the crystal $b$-axis and the current flow direction. Angle $\theta$ is defined as the angle between the field direction and the $b$-axis. In the AFM ordered region ($T$ $<$ $T_N$, Fig. \ref{MR} (a)), at $\theta$=90$^\circ$, MR first increases rapidly with field, as expected for compounds in the AFM state. Beyond a certain field, MR decreases quickly, becomes negative, and finally saturates in the high fields. With the field rotated towards the $b$-axis, both the positive and negative MR effects are suppressed, and the evolution from the positive to negative MR with field becomes progressively smooth. With increasing temperature (Fig. \ref{MR} (b)), qualitatively similar MR was seen although the positive MR now becomes very weak. At $T$=75 K, The MR becomes negligibly small and only positive MR and an inflection point are seen for $H$$\parallel$$b$-axis. Figs. \ref{MR} (d-f) plot the MR at three angles under various temperatures. Similar behavior is seen for all these angles although the size of the MR gets larger as the angle increases.

As is known, for a single-ion Kondo system, the MR curves measured under different temperatures can be scaled by the Schlottmann's relation, where $T^*$ is a scaling parameter related to magnetic correlation types\cite{Schlottmann}. Figs. \ref{MR} (g-f) show the Schlottmann's scaling for three different angles for data measured above the AFM transition temperature $T_N$. As seen, for H $\parallel$ $b$, Schlottmann's scaling is violated while for $H$ $\parallel$ $ac$ and for the intermediate angle ($\theta$= 50$^\circ$), all isothermal MR curves collapse onto a single curve, in good agreement with Schlottmann's theory. The best scaling yields a negative $T^*$=-1.8 K, implying the presence of ferromagnetic correlations in this antiferromagnetic system. Similar Schlottmann's scaling has also been observed in other heavy-fermion systems such as UBe$_{1.3}$\cite{Andraka}, YbPtSn\cite{Pietri}, and CeNiGe$_3$\cite{Sato}. The origin of the anisotropic violation of the Schlottmann's scaling in CeBi$_2$ is unclear to us at present.

In Fig. \ref{highfield}, high-field MR measured up to 35 T for two pieces of samples at $T$ = 4.2 K is shown. Note that these two samples show larger MR than that shown in Fig. \ref{MR} at the same field strengths, due to higher sample purity. Although the MR behaviors are qualitatively similar to those shown in Fig. \ref{MR}, the quasi-linear MR in the high fields is intriguing and deserves more investigations in due course\cite{PdSn4,RhSn4}.

Finally, let us turn to consider the origin of the negative MR seen in this magnetic Kondo system. As shown in Fig. \ref{comparefig} (more pronounced in the panel (b)), the initial field where the positive MR start to change to the negative one coincides with the field for the meta-magnetic phase transition, suggesting the possible symbiosis between these two phenomena. Besides, the MR saturates at the high field that may also be associated with the ferromagnetic-like saturation seen in the magnetization curves. Therefore, it is reasonable to attribute the intriguing MR in this system to its intricate magnetism, rather than the orbital effects or relativistic chiral anomaly physics\cite{OngNa3Bi,OngCd3As2}.

\section{Conclusion}

In summary, we studied the detailed transport and thermodynamic properties of the antiferromagnetic Kondo system CeBi$_2$ that displays an antiferromagnetic transition with the N\'{e}el temperature $T_N$ around 3.3 K. By analyzing the specific heat data, we found the Sommerfeld coefficient $\gamma$ over 200 mJ/mol K$^2$ in its paramagnetic state and a Kondo temperature of an order of $\sim$ 2 K. We presented the evidence that the large negative magnetoresistance observed in the wide temperature and field-angle ranges in this system is associated with its complex magnetism. CeBi$_2$ therefore represents an interesting system to formulate the theoretical understanding of the exotic consequences from the interplay between local $f$-moments and itinerant conduction electrons.

\begin{acknowledgments}
The authors would like to thank Nigel Hussey, Xin Lu, Pabitra Biswas, C. M. J. Andrew for the fruitful discussion. This work is sponsored by the National Key Basic Research Program of China (Grant No. 2014CB648400), by National Natural Science Foundation of China (Grant No. 11474080, No. U1732162, No. 11704047, No. 11374043) and by Natural Science Foundation of Jiangsu Educational Department (Grant No. 15KJA430001) and six-talent peak of Jiangsu Province (Grants No. 2012-XCL-036). We acknowledge the support of the HFML, member of the European Magnetic Field Laboratory (EMFL). X. X. would also like to acknowledge the financial support from an open program from Wuhan National High Magnetic Field Center (2015KF15).

W. Zhou, C. Q. Xu contributed equally to this work.
\end{acknowledgments}

\appendix

\end{document}